\documentclass[twocolumn]{ijns}
\usepackage{graphicx,epsfig,color}
\usepackage[normal]{subfigure}
\usepackage{amsmath,amsthm,amssymb}
\usepackage{floatflt}  
\usepackage{setspace}
\usepackage{rotating,algorithm,algorithmic}
\usepackage{multirow}
\usepackage{titlecaps}
\usepackage{etoolbox}
\usepackage{hyperref}
\apptocmd{\thebibliography}{\raggedright}{}{}

\def\markboth#1#2{\def\leftmark{#1}\def\rightmark{#2}}

\UseRawInputEncoding

\title{\huge\bf Classifying Malware Images with Convolutional Neural Network Models}
\author{Ahmed Bensaoud, Nawaf Abudawaood, and Jugal Kalita\\
{\normalsize Department of Computer Science, University of Colorado Colorado Springs, USA}\\
{\normalsize (Email: abensaou@uccs.edu)}\\
}

\date
\newpage
\begin{document}
\maketitle

\begin{abstract}
Due to increasing threats from malicious software (malware) in both number and complexity, researchers have developed approaches to automatic detection and classification of malware, instead of analyzing methods for malware files manually in a time-consuming effort. At the same time, malware authors have developed techniques to evade signature-based detection techniques used by antivirus companies. Most recently, deep learning is being used in malware classification to solve this issue. In this paper, we use several convolutional neural network (CNN) models for static malware classification. In particular, we use six deep learning models, three of which are past winners of the ImageNet Large-Scale Visual Recognition Challenge. The other three models are CNN-SVM, GRU-SVM and MLP-SVM, which enhance neural models with support vector machines (SVM). We perform experiments using the Malimg dataset, which has malware images that were converted from Portable Executable malware binaries. The dataset is divided into 25 malware families. Comparisons show that the Inception V3  model achieves a test accuracy of 99.24\%, which is better than the accuracy of 98.52\% achieved by the current state-of-the-art system called the M-CNN model.
\vspace*{0.1cm}
~
\end{abstract}

\section{Introduction}

Internet connectivity is an essential infrastructure for business organizations, banking institutions, universities, and governments, and is growing exponentially.  This growth is threatened by attackers with malicious codes and network threats\cite{Tayal}. The execution of malware forces a computer to perform operations that are not normal, and may harm a victim's computer systems. The amount of malware in circulation has been increasing rapidly in the recent years, and malware has affected computer systems all over the world\cite{kolosnjaji2016deep}. Thousands of malware files are being created daily. Fig. 1 presents annual statistics of malware attacks over the last 10 years, showing that the total number of malware  in circulation has increased to more than 900 million in 2019, which is a 2000\% increase compared to the number of malware in the year 2010\cite{StaReport}. 

The cost of malware infection can run into millions of dollars for each incident inflicted upon small and medium sized businesses\cite{IRSReport}. Routing protocols alone are not sufficient to detect malware\cite{ zareapoor2014establishing}. As a result, researchers and anti-virus vendors employ machine learning to detect and classify malicious software. A large number of studies have focused on malware binary since binaries are normally used to infect computers. Malware is analyzed based on static as well as dynamic analysis. While static analysis extracts malware features that can be used to detect or classify malware employing machine learning, dynamic analysis analyzes malware behavior as it is executed in a controlled environment like Cuckoo Sandbox \cite{guarnieri2013cuckoo}, which is open source, available on GitHub.

\begin{figure} [hbt!]
\begin{center}
\fbox{\includegraphics[width=0.95\linewidth,height=0.3\textheight]{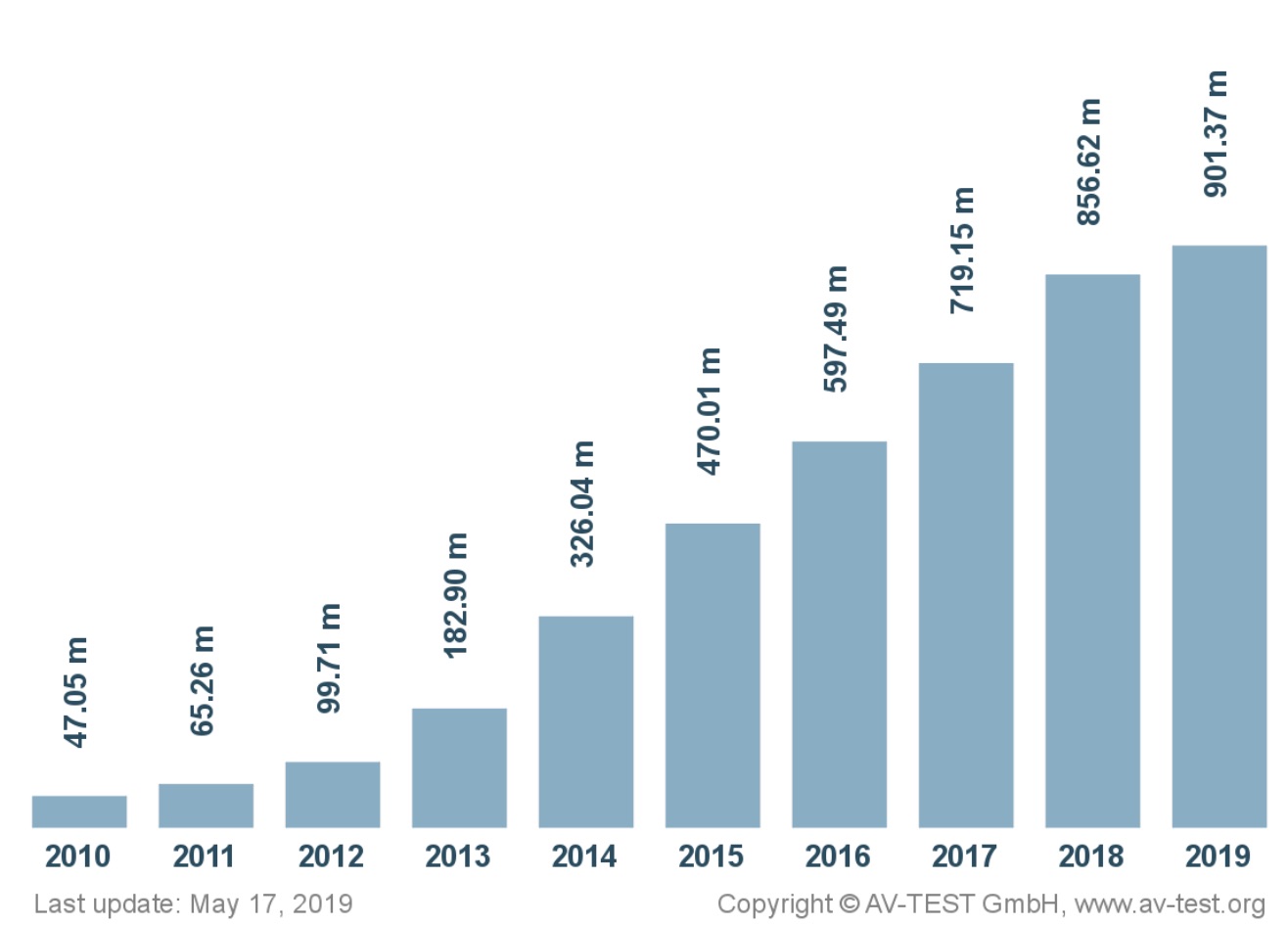}}
	\caption{Number of worldwide malware attacks for the last ten years.}
	\end{center}
\end{figure}

Various traditional machine learning approaches such as support vector machine\cite{keerthi2002convergence}, k-nearest neighbors \cite{gianfelici2008nearest}, random forests\cite{liaw2002classification}, naive bayes\cite{domingos1997optimality} and decision tree\cite{quinlan1986induction} have been used to detect and classify known malware. In particular, Nataraj et al.\cite{ nataraj2011malware} proposed a method for visualizing and classifying malware using image processing methods, which first converts malware binaries to grayscale images. Techniques from computer vision, particularly for image classification can be used to obtain high accuracies.

Researchers have classified malware using CNN models, initially used for image classification \cite{sewak2018comparison}. It is obvious that in order to use such an approach, the malware binary must first be converted to an ``image''. The ANN models used include simple multilayer perceptron, and a mix of GRU-based RNNs and CNNs. Kalash et al.\cite{kalash2018malware} used a CNN model called M-CNN, based on a well-known image classification architecture called VGG-16 \cite{simonyan2014very}. Methods have also replaced the last layer of an artificial neural network with an SVM classifier\cite{niu2012novel}.

In this paper, we compare the performance of several CNN-based models which had achieved state-of-the-art results for malware image classification with the CNN-mixed models used by Agarap and Pepito \cite{agarap2017towards}. The CNN models we choose have performed well in the large-scale image classification contest called ILSVRC\cite{russakovsky2015imagenet}, within the last few years.

The paper is organized in the following way. In the next section, we briefly review related work. Section 3 describes the methodology used to classify malware. Section 4 discusses experimental results. Lastly, Section 5 concludes the paper and discusses plans for future work. 

\section{Related work}

Below, we discuss research effects that primarily convert malware binaries to images before classifying them. Approaches based on traditional machine learning depend on manual feature extraction. Deep learning can extract useful features automatically by avoiding manual feature extraction.

\subsection{Methods based on traditional machine learning}

Grayscale images can be extracted from the raw malware executable files showing features of malware \cite{nataraj2011comparative}\cite{nataraj2011malware}\cite{kosmidis2017machine}. Such images enable analysis of malware by extracting visual features. Nataraj et al. \cite{nataraj2011malware} were the first to explore the use of byte plot visualization as grayscale images for automatic malware classification. They used a malware image dataset consisting of 9,342 malware samples belonging to 25 different classes. They extracted GIST\cite{torralba2003context} features from the grayscale images and classified them using K-nearest neighbor classification with Euclidean distance as metric. Their approach had high computational overhead. Mirza et al. extracted features from malware files and combined decision trees, support vector machines and boosting to detect malware\cite{mirza2018cloudintell}. Zhang et al. proposed a static analysis technique based on n-grams of opcodes to classify ransomware families \cite{zhang2019classification}.
Makandar and Patrot\cite{makandar2017malware} used multi-class support vector machine malware classification with malware input as images. They used wavelet transform to build effective texture based feature vectors from the malware images. This reduced the dimensionality of the feature vector and the time complexity.
\subsection{Methods based on deep learning}
Several studies on malware classification have been performed using CNN architectures. Cui et al.\cite{cui2018detection} detected code variants that are malicious after converting to grayscale images and using a simple CNN model. Kalash et al.\cite{kalash2018malware} classified malware images by converting malware files into grayscale images, using two different datasets, Malimg \cite{ nataraj2011malware} and Microsoft Malware \cite{ronen2018microsoft}. They obtained 98.52\% and 99.97\% accuracies, respectively. Yue\cite{yue2017imbalanced} proposed a weighted softmax loss for CNNs for imbalanced malware image classification, and achieved satisfactory classification results. Gilbert. et al.\cite {gibert2019using} built a model consisting of three convolutional layers with one fully connected layer and tested on two datasets, Microsoft Malware Classification Challenge dataset and Malimg dataset. Seonhee et al. \cite{ seok2016visualized} proposed a malware classification model using a CNN that classified malware images. Their experiments were divided into two sets. The first set of experiments classified malware into 9 families and obtained accuracies of 96.2\% and 98.4\% considering the top-1 and top-2 ranked results. The second set of experiments classified malware into 27 families and obtained 82.9\% and 89\% top-1 and top-2 accuracies. Tobiyama et al.\cite{tobiyama2016malware} proposed a malware process detection method by training a recurrent neural network (RNN) to extract features of process behavior, and then training a CNN to classify features extracted by the trained RNN. Vinayakumar et al. proposed a deep learning model based on CNN and LSTM for malware family categorization. Experiments showed an accuracy of 96.3\% on the Malimg dataset \cite {vinayakumar2019robust}. Su et al.\cite{su2018lightweight} created one-channel grayscale images from executable binaries in two families, and classified them into their related families using a light-weight Convolutional Neural Network. They achieved a accuracies of 94.0\% and 81.8\% for malware and goodware, respectively.
\section{Methodology}

In this paper, we use six CNN models for malware
classification, considering malware binaries as images.

\subsection{Malware Binaries}

The malware binaries we use are in Portable Executable (PE) form. Generally, PE files are programs that have file name extensions such as .bin, .dll and .exe. PE files are usually recognized through their components, which are called .tex, .rdata, .data and .rsrc. The first component, called .text, is the code section, containing the program's instructions. .rdata is the part that contains read only data, and .data is the part that contains data that can be modified, and .rsrc is the final component that stands for resources used by the malware.

Malicious data binaries can be converted 8 bits at a time to pixels in a grayscale image, consisting of textural patterns. In Fig. 2, we see the sections of a malware binary showing different textures, when seen as an image\cite{ nataraj2011malware}. Based on these patterns, we can classify malware. In this paper, we use the Malimg dataset\cite{ nataraj2011malware} which is a set of grayscale images corresponding to malware binaries saved in .jpg format. Some examples of malware families are shown in Fig. 3.

\begin{figure} [hbt!]
\centering
\includegraphics[width=0.7\linewidth,height=0.3\textheight]{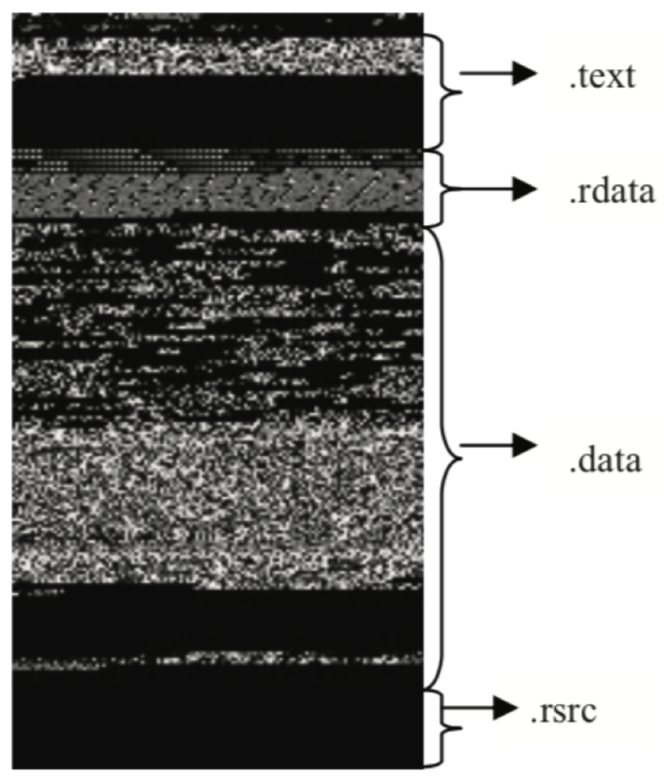}

	\caption{Portable Executable file represented as an image.}
\end{figure}

\begin{figure} [hbt!]
\centering
\includegraphics[width=0.6\linewidth,height=0.4\textheight]{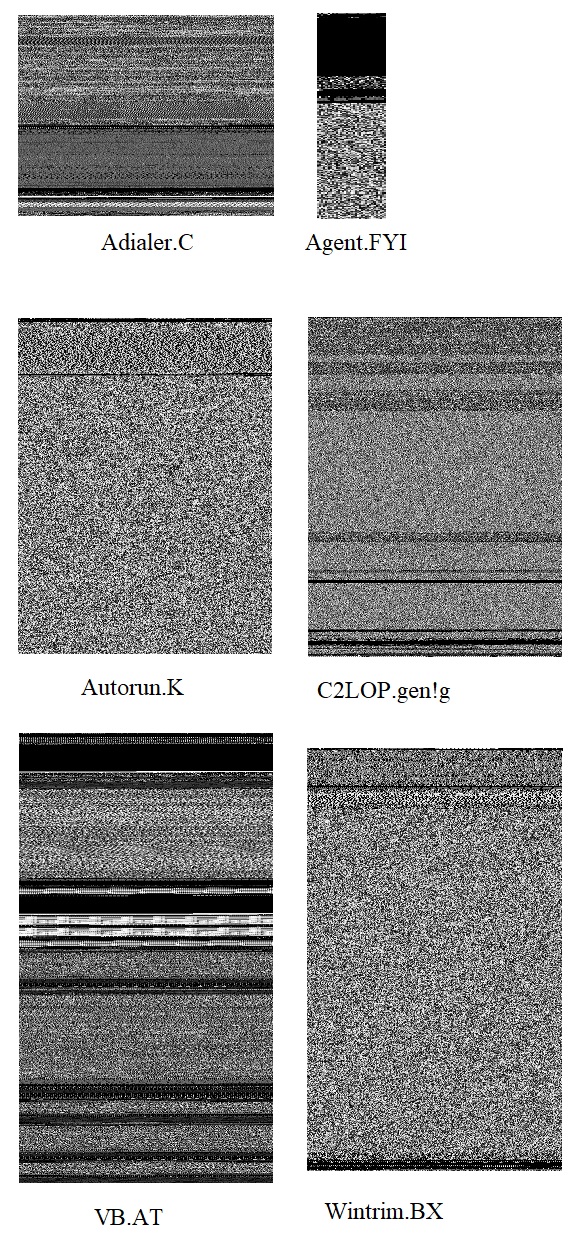}
	\caption{Sample images of malware belonging to different families.}
\end{figure}

\subsection{Malware as image}
Researchers and practitioners can understand malware better by visualizing malware binaries as images since the patterns within such images become clearly visible. Finding patterns within images can be performed well by deep learning \cite{gu2018recent}. The most important patterns of features in the malware images can be used to identify the malware families also. Images for a specific malware family have similar patterns, allowing a deep learning model to recognize important patterns using automatic extraction of features. In particular, CNN models are good at classifying images because they can extract relevant features within an image by subsampling through convolutions, pooling and other computations. In this case, CNNs look for the most relevant features within an image from a specific malware family for the purpose of classification \cite{cui2018detection}. Malware binaries can be translated into images using an algorithm that converts a binary PE file into a sequence of $8$ bit vectors or hexadecimal values. An $8$ bit vector can be represented in the range $00000000$ $(0)$ to $11111111$ $(255)$. Each $8$ bit vector represents a number, and can be converted into pixel in a malware image, as shown in Fig. 3. Images obtained from different malware families have different characteristics \cite{kalash2018malware}.

\begin{figure} [hbt!]
\fbox{\includegraphics[width=0.9\linewidth,height=0.1\textheight]{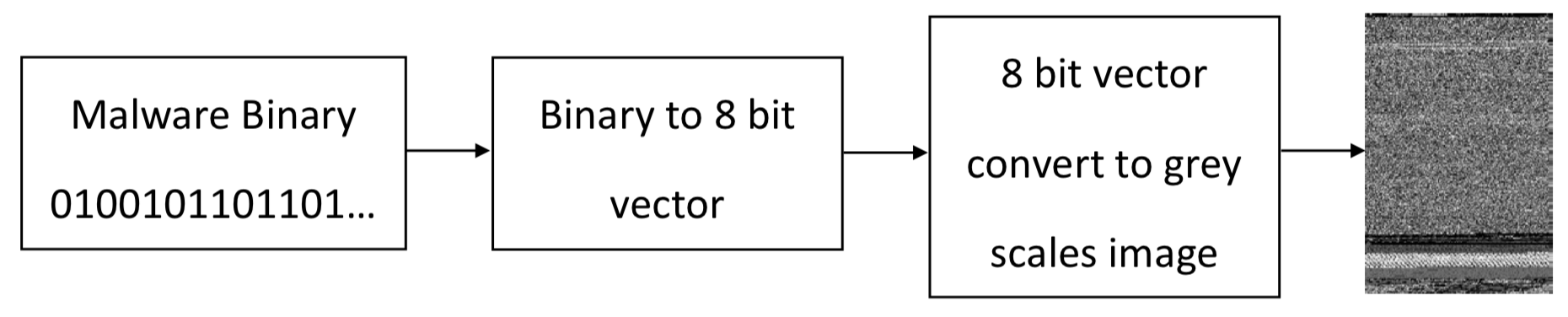}}
\caption{Converting malware binary to an image.}
\end{figure}

\subsection{Problem Statement}

The problem that we solve in this paper is classification of malware object code into malware families. We have 9,342 malware samples given in the form of images obtained from their object code. There are 25 malware families, with the biggest family containing 2,950 samples and the smallest containing 81 samples. We classify these images using deep learning models that have performed well in image classification. 

\subsection{Motivation and Approach}
CNNs have performed well for classification in a variety of domains including object recognition \cite{kavukcuoglu2010learning}, image classification \cite{krizhevsky2012imagenet}, and video classification \cite{karpathy2014large}. CNNs have shown superior performance compared to traditional learning algorithms, especially in tasks such as image classification. Since we represent malware object code as images, we classify malware based on their corresponding images using CNN models. Malware images are classified into families by extracting patterns within them, because binary image files generated from a malware family are likely to produce similar images. Feature extraction allows image classification models to recognize patterns based on pixel distribution in an image. Before CNNs, features were extracted manually, and it was one of the biggest challenges in image classification. The ImageNet Large-Scale Visual Recognition Challenge (ILSVRC)\cite{russakovsky2015imagenet} has led to sophisticated CNN-based classification models that have achieved excellent results, demonstrating that the models are likely to perform well in static analysis of malware. 

In this paper, we compare the performance of several CNNs-based models for classification of malware binaries that have been converted to images. In particular, we compare the performance of several well-known CNNs-based deep learning models from the ILSVRC competitions and a few additional CNN and CNN-mixed models to classify malware images, models that automatically extract features based on the static analysis approach. These models are publicly available. 

\subsection{CNN Models Used}

The experimental work of this paper is to run six deep learning models to classify malware images to detect malware. These models are briefly described below.

\subsubsection{VGG16}
The first model we use is called VGG-Net16 \cite{simonyan2014very}, which was the winner of ILSVRC in 2014. Its contribution was in increasing the depth using 3x3 convolution filters that are small, allowing them to increase the number of layers from 16 to 19. The depth of the representation was very helpful in increasing the accuracy of image classification. On the ImageNet dataset, the VGG model outperformed many complicated models, signifying the importance of the depth.

\begin{figure} [hbt!]
\centering

\fbox{\includegraphics[width=0.95\linewidth,height=0.2\textheight]{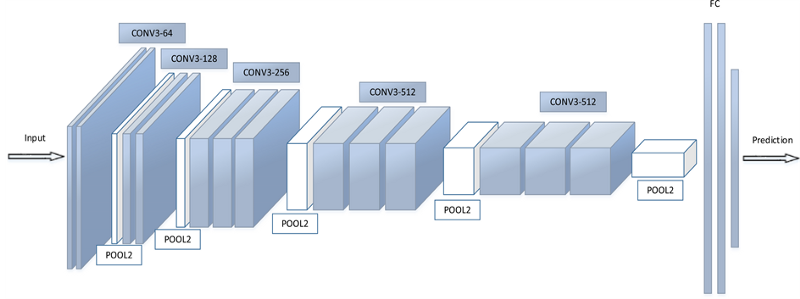}}
	\caption{VGG-16 model architecture\cite{vgginception}.}
\end{figure}

\subsubsection{Inception V3}

The Inception V3 model contains 42 layers, and is an improvement over the GoogleNet Inception V1 model that was the winner of ILSVRC in 2015\cite{szegedy2016rethinking}. The Inception V3 model architecture starts with a 5x Inception module A, 4x Inception module B, 2x Inception module C, and 2x grid size reduction; one of the grid size reductions is done with some modification, and the second one is applied without any modification. An auxiliary classifier is also applied as an extra layer to help improve the results.

\begin{figure} [hbt!]
\centering
\fbox{\includegraphics[width=0.95\linewidth,height=0.3\textheight]{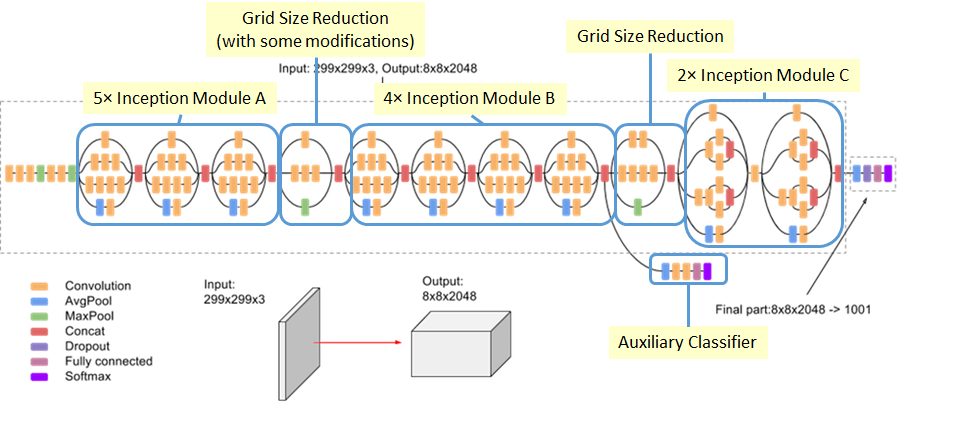}}
	\caption{Inception V3 model architecture\cite{sikinception}.}
\end{figure}

\subsubsection{ResNet50}
The third model we use is called Residual Networks (ResNet50)\cite{he2016deep}. ResNet50 was the winner of ILSVRC in 2016. The novel technique that this model introduced provides extra connections between non-contiguous convolutional layers, using shortcut connections. This technique allowed the model to skip through layers to deal with vanishing gradients in order to achieve lower loss and better results. The network had 152 layers, an impressive 8 times deeper than a comparable VGG network. This is an improvement over the VGG16 model with Faster R-CNN, producing an improvement of 28\% in accurcy in image classification. The architecture of the original ResNet50 is illustrated in Fig. 7.

\begin{center}
\begin{figure} [hbt!]
\centering
\fbox{\includegraphics[width=0.95\linewidth,height=0.3\textheight]{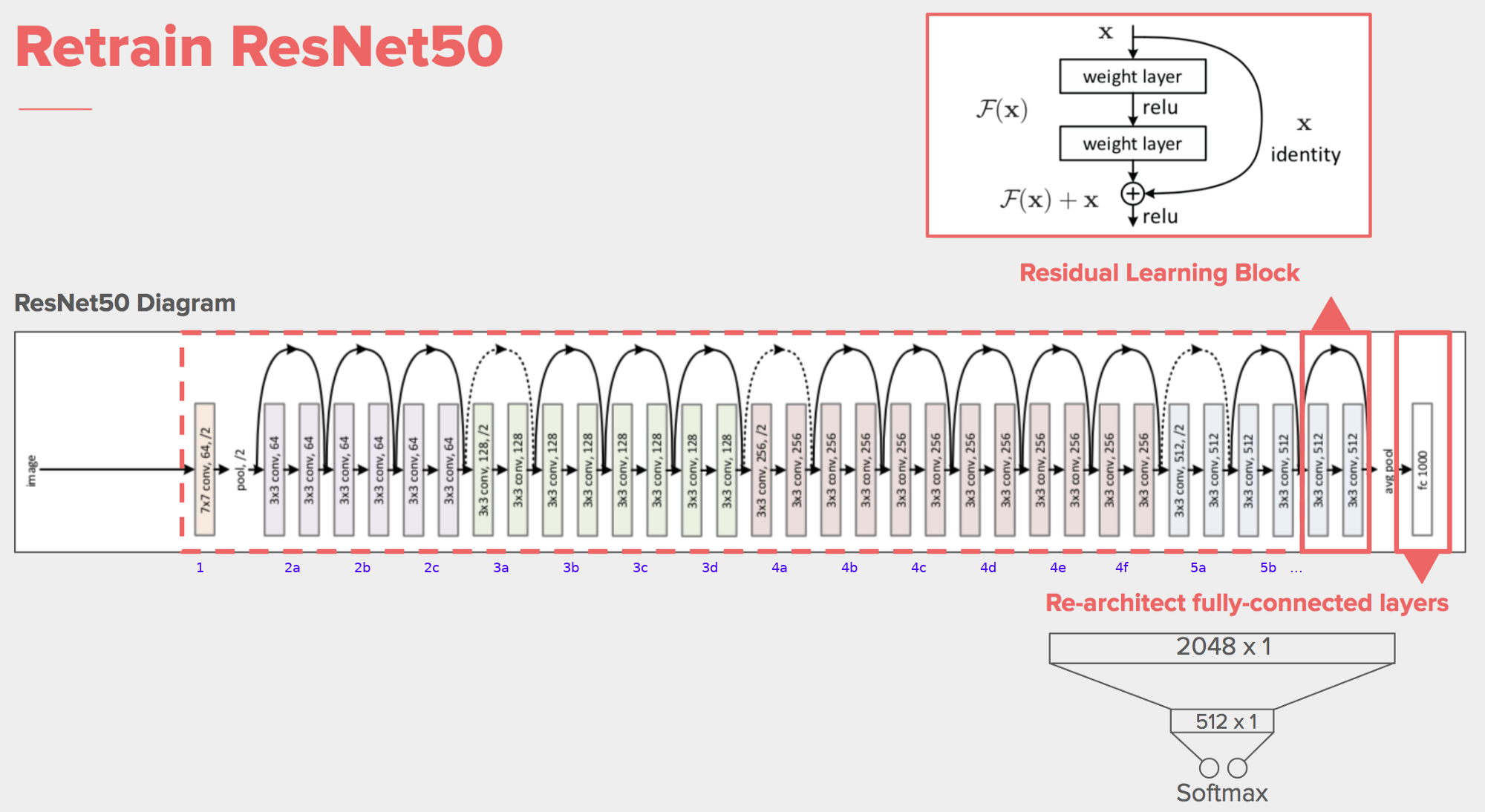}}
	\caption{ResNet50 model architecture\cite{adriaresnet50}.}
\end{figure}
\end{center}

\subsubsection{CNN-SVM model}

For classification, deep learning models usually use the softmax activation function as the top layer for prediction and minimization of cross-entropy loss. Tang\cite{tang2013deep} replaced the softmax layer with a linear SVM and applied it on MNIST and CIFAR-10 datasets, and the ICML 2013 Representation Learning Workshop’s face expression recognition challenge. The SVM is a linear maximum margin classifier. CNN-SVM allowed for extraction of features for input images with a linear SVM\cite{fanany2017handwriting}. Agarap and Pepito\cite{agarap2017towards} applied CNN-SVM\cite{tang2013deep} on Malimg and achieved 77.22\% accuracy.

\begin{center}
\begin{figure} [hbt!]
\fbox{\includegraphics[width=0.95\linewidth,height=0.2\textheight]{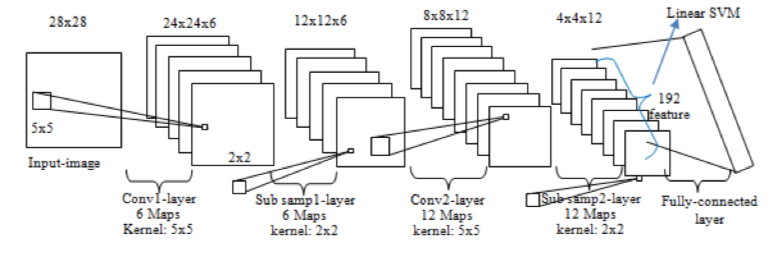}}
	\caption{Architecture of CNN-SVM\cite{Darmatasia2017HandwritingRO}.}
\end{figure}
\end{center}

\subsubsection{GRU-SVM model}

Agarap and Pepito\cite{agarap2017towards} modified the architecture of a Gated Recurrent Unit (GRU) RNN by using SVM as its final output layer for use in a binary, non-probabilistic classification task (see Fig 8). They used GRU-SVM on the Malimg dataset and achieved 84.92\% accuracy.

\begin{center}
\begin{figure} [hbt!]

\fbox{\includegraphics[width=0.9\linewidth,height=0.2\textheight]{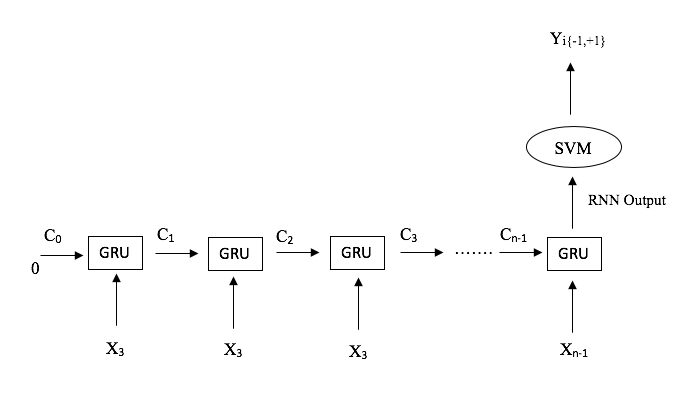}}

	\caption{GRU-SVM architecture model, with n GRU cells and SVM for the classification function\cite{agarap2017neural}. }
\end{figure}
\end{center}

\subsubsection{MLP-SVM model}
Bellili et al.\cite{bellili2001hybrid} proposed MLP-SVM for handwritten digit recognition. MLP-SVM is a model that combines both SVM and Multilayer Perceptrons for the classification of binary image. Multilayer Perceptrons are a fully connected network that allows for the inputs to get classified using input features. The MLP-SVM is a hybrid model that runs the MLP and SVM classifiers in parallel. The MLP-SVM model was used by Agarap and Pepito\cite{agarap2017towards} on the Malimg dataset with 80.46\% accuracy.

\begin{center}
\begin{figure} [hbt!]
\fbox{\includegraphics[width=0.95\linewidth,height=0.3\textheight]{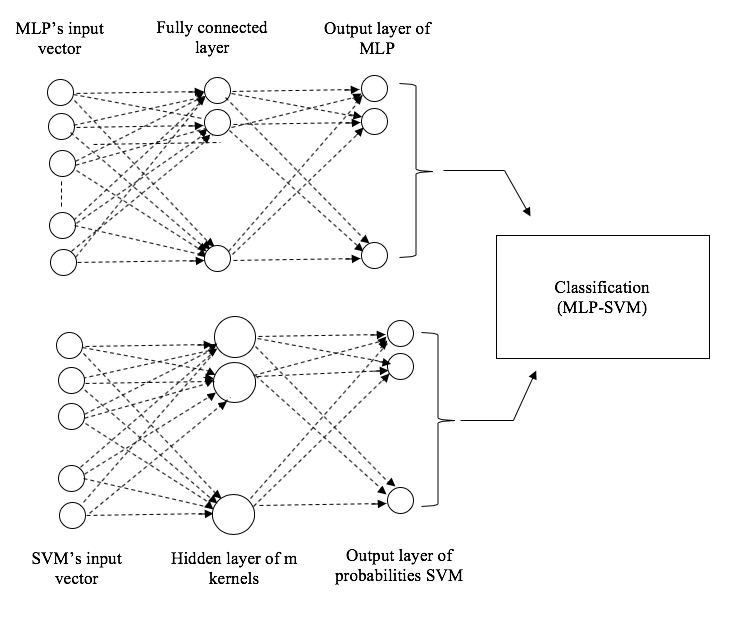}}

	\caption{MLP-SVM architecture model\cite{tra2018diagnosis}.}
\end{figure}
\end{center}

\subsection{\textbf{Dataset}}

There are a few malware datasets available for academic research. One of the these datasets is Malimg\cite{nataraj2011malware}. The dataset contains 9,342 malware images, classified into 25 malware families. The widths and lengths of the malware images vary. The images have been created from various malware families such as Dialer, Backdoor, Worm, Worm-AutoIT, Trojan, Trojan-Downloader, Rouge and PWS. All malware images are PE files that were first converted to an 8-bit vector binary, and then to images. The malware image sizes were modified, so that they can be input to a CNN model. The family breakdown for the Malimg dataset is shown in Table 1.

\begin{table} [hbt!]
 \caption{25 malware families (classes) and the number of samples in each family.}
\begin{tabular}{|c|c|c|}
\hline
\textbf{Malware Family} & \textbf{Samples} & \textbf{Malware kind} \\ \hline
Adialer.C & 123 & Dialer\\ \hline
Agent.FYI & 117 & Backdoor\\ \hline
Allaple.A & 2950& Worm \\ \hline
Allaple.L     & 1592& Worm\\ \hline
Alueron.gen!J & 199& Trojan\\ \hline
Autorun.K     & 107& Worm AutoIT\\ \hline
C2LOP.gen!g   & 201& Trojan\\ \hline
C2LOP.p        & 147& Trojan\\ \hline
Dialplatform.B & 178& Dialer\\ \hline
Donoto.A       & 163& Trojan Downloader\\ \hline
Fakerean       & 382& Rouge\\ \hline
Instaccess     & 432& Dialer\\ \hline
Lolyada.AA1    & 214& PWS\\ \hline
Lolyada.AA2    & 185& PWS\\ \hline
Lolyada.AA3    & 124& PWS\\ \hline
Lolyada.AT    & 160& PWS\\ \hline
Malex.gen!J   & 137& Trojan\\ \hline
Obfuscator.AD  & 143& Trojan Downloader\\ \hline
RBot!gen      & 159& Backdoor\\ \hline
Skintrim.N    & 81& Trojan\\ \hline
Swizzor.gen!E   & 129& Trojan Downloader\\ \hline
Swizzor.gen!I   & 133& Trojan Downloader\\ \hline
VB.AT       & 409& Worm\\ \hline
Wintrim.BX  & 98& Trojan Downloader\\ \hline
Yuner.A    & 801& Worm\\ \hline
\end{tabular}
\end{table}

\begin{table*} [!hbpt]
\caption{Prediction accuracies of the six tested models.}
\centering
\begin{tabular}{|p{0.9cm}|p{0.9cm}|p{0.9cm}|p{0.9cm}|p{0.9cm}|p{0.9cm}|p{0.9cm}|}
\hline
\multicolumn{1}{|c|}{\multirow{2}{*}{\textbf{Family}}} & \multicolumn{1}{l|}{\textbf{CNN-SVM}} & \multicolumn{1}{l|}{\textbf{GRU-SVM}} & \multicolumn{1}{l|}{\textbf{MLP-SVM}} & \multicolumn{1}{l|}{\textbf{Inception V3}}  & \multicolumn{1}{l|}{\textbf{ResNet 50}}  & \multicolumn{1}{l|}{\textbf{VGG16}}\\ \cline{2-7} 
\multicolumn{1}{|c|}{}                        & \multicolumn{6}{c|}{\textbf{Prediction Accuracy}}                                                                                                                      \\ \hline
\multicolumn{1}{|l|}{Adialer.C}               & \multicolumn{1}{l|}{\textbf{99.80\%}}        & \multicolumn{1}{l|}{99.15\%}        & \multicolumn{1}{l|}{99.51\%}        & \multicolumn{1}{l|}{99.40\%}    & \multicolumn{1}{l|}{23.18\%}& \multicolumn{1}{l|}{13.62\%}        \\ \hline
\multicolumn{1}{|l|}{Agent.FYI}               & \multicolumn{1}{l|}{95.12\%}        & \multicolumn{1}{l|}{95.86\%}        & \multicolumn{1}{l|}{94.87\%}        & \multicolumn{1}{l|}{\textbf{99.50\%}}    & \multicolumn{1}{l|}{25.41\%}& \multicolumn{1}{l|}{14.81\%}         \\ \hline
\multicolumn{1}{|l|}{Allaple.A}               & \multicolumn{1}{l|}{94.98\%}        & \multicolumn{1}{l|}{97.71\%}        & \multicolumn{1}{l|}{94.32\%}        & \multicolumn{1}{l|}{\textbf{99.72\%}}    & \multicolumn{1}{l|}{26.94\%}& \multicolumn{1}{l|}{14.47\%}        \\ \hline
\multicolumn{1}{|l|}{Allaple.L}               & \multicolumn{1}{l|}{99.10\%}        & \multicolumn{1}{l|}{95.35\%}        & \multicolumn{1}{l|}{95.48\%}        & \multicolumn{1}{l|}{\textbf{99.73\%}}    & \multicolumn{1}{l|}{21.52\%}& \multicolumn{1}{l|}{15.53\%}        \\ \hline
\multicolumn{1}{|l|}{Alueron.gen!J}           & \multicolumn{1}{l|}{96.42\%}        & \multicolumn{1}{l|}{97.57\%}        & \multicolumn{1}{l|}{93.20\%}        & \multicolumn{1}{l|}{\textbf{99.48\%}}    & \multicolumn{1}{l|}{21.37\%}& \multicolumn{1}{l|}{15.93\%}               \\ \hline
\multicolumn{1}{|l|}{Autorun.K}               & \multicolumn{1}{l|}{92.99\%}        & \multicolumn{1}{l|}{93.38\%}        & \multicolumn{1}{l|}{96.68\%}        & \multicolumn{1}{l|}{\textbf{99.06\%}}    & \multicolumn{1}{l|}{23.27\%}& \multicolumn{1}{l|}{14.38\%}              \\ \hline
\multicolumn{1}{|l|}{C2LOP.gen!g}             & \multicolumn{1}{l|}{94.75\%}        & \multicolumn{1}{l|}{93.70\%}        & \multicolumn{1}{l|}{94.49\%}        & \multicolumn{1}{l|}{\textbf{98.42\%}}    & \multicolumn{1}{l|}{28.87\%}& \multicolumn{1}{l|}{13.78\%}               \\ \hline
\multicolumn{1}{|l|}{C2LOP.P}                 & \multicolumn{1}{l|}{97.11\%}        & \multicolumn{1}{l|}{93.45\%}        & \multicolumn{1}{l|}{95.43\%}        & \multicolumn{1}{l|}{\textbf{99.67\%}}    & \multicolumn{1}{l|}{27.48\%}& \multicolumn{1}{l|}{14.96\%}              \\ \hline
\multicolumn{1}{|l|}{Dialplatform.B.}         & \multicolumn{1}{l|}{95.34\%}        & \multicolumn{1}{l|}{94.85\%}        & \multicolumn{1}{l|}{96.17\%}        & \multicolumn{1}{l|}{\textbf{99.86\%}}    & \multicolumn{1}{l|}{23.84\%}& \multicolumn{1}{l|}{14.78\%}              \\ \hline
\multicolumn{1}{|l|}{Dontovo.A}               & \multicolumn{1}{l|}{97.53\%}        & \multicolumn{1}{l|}{89.81\%}        & \multicolumn{1}{l|}{93.44\%}        & \multicolumn{1}{l|}{\textbf{98.25\%}}    & \multicolumn{1}{l|}{29.76\%}& \multicolumn{1}{l|}{15.03\%}             \\ \hline
\multicolumn{1}{|l|}{Fakerean}                & \multicolumn{1}{l|}{98.46\%}        & \multicolumn{1}{l|}{92.11\%}        & \multicolumn{1}{l|}{93.11\%}        & \multicolumn{1}{l|}{\textbf{98.91\%}}    & \multicolumn{1}{l|}{26.29\%}& \multicolumn{1}{l|}{12.45\%}                \\ \hline
\multicolumn{1}{|l|}{Instantaccess.}          & \multicolumn{1}{l|}{93.17\%}        & \multicolumn{1}{l|}{96.75\%}        & \multicolumn{1}{l|}{96.63\%}        & \multicolumn{1}{l|}{\textbf{98.24\%}}    & \multicolumn{1}{l|}{30.15\%}& \multicolumn{1}{l|}{13.11\%}            \\ \hline
\multicolumn{1}{|l|}{Lolyda.AA1}              & \multicolumn{1}{l|}{91.30\%}        & \multicolumn{1}{l|}{94.09\%}        & \multicolumn{1}{l|}{93.97\%}        & \multicolumn{1}{l|}{\textbf{99.40\%}}    & \multicolumn{1}{l|}{23.79\%}& \multicolumn{1}{l|}{14.26\%}             \\ \hline
\multicolumn{1}{|l|}{Lolyda.AA2}              & \multicolumn{1}{l|}{89.10\%}        & \multicolumn{1}{l|}{94.36\%}        & \multicolumn{1}{l|}{91.64\%}        & \multicolumn{1}{l|}{\textbf{99.34\%}}    & \multicolumn{1}{l|}{28.32\%}& \multicolumn{1}{l|}{13.80\%}              \\ \hline
\multicolumn{1}{|l|}{Lolyda.AA3}              & \multicolumn{1}{l|}{87.44\%}        & \multicolumn{1}{l|}{90.61\%}        & \multicolumn{1}{l|}{94.13\%}        & \multicolumn{1}{l|}{\textbf{97.39\%}}    & \multicolumn{1}{l|}{29.59\%}& \multicolumn{1}{l|}{13.85\%}              \\ \hline
\multicolumn{1}{|l|}{Lolyda.AT}               & \multicolumn{1}{l|}{81.31\%}        & \multicolumn{1}{l|}{92.51\%}        & \multicolumn{1}{l|}{90.28\%}        & \multicolumn{1}{l|}{\textbf{99.86\%}}    & \multicolumn{1}{l|}{31.67\%}& \multicolumn{1}{l|}{13.99\%}          \\ \hline
\multicolumn{1}{|l|}{Malex.gen!J}             & \multicolumn{1}{l|}{88.79\%}        & \multicolumn{1}{l|}{94.99\%}        & \multicolumn{1}{l|}{94.61\%}        & \multicolumn{1}{l|}{\textbf{99.31\%}}    & \multicolumn{1}{l|}{25.39\%}& \multicolumn{1}{l|}{15.22\%}              \\ \hline
\multicolumn{1}{|l|}{Obfuscator.AD.}          & \multicolumn{1}{l|}{86.57\%}        & \multicolumn{1}{l|}{94.76\%}        & \multicolumn{1}{l|}{96.74\%}        & \multicolumn{1}{l|}{\textbf{99.50\%}}    & \multicolumn{1}{l|}{21.84\%}& \multicolumn{1}{l|}{12.64\%}           \\ \hline
\multicolumn{1}{|l|}{Rbot!gen.}               & \multicolumn{1}{l|}{87.60\%}        & \multicolumn{1}{l|}{93.39\%}        & \multicolumn{1}{l|}{97.19\%}        & \multicolumn{1}{l|}{\textbf{98.81\%}}    & \multicolumn{1}{l|}{32.49\%}& \multicolumn{1}{l|}{14.45\%}             \\ \hline
\multicolumn{1}{|l|}{Skintrim.N}              & \multicolumn{1}{l|}{96.16\%}        & \multicolumn{1}{l|}{84.10\%}        & \multicolumn{1}{l|}{87.21\%}        & \multicolumn{1}{l|}{\textbf{99.55\%}}    & \multicolumn{1}{l|}{34.81\%}& \multicolumn{1}{l|}{15.84\%}              \\ \hline
\multicolumn{1}{|l|}{Swizzor.gen!E.}          & \multicolumn{1}{l|}{82.45\%}        & \multicolumn{1}{l|}{96.72\%}        & \multicolumn{1}{l|}{98.54\%}        & \multicolumn{1}{l|}{\textbf{99.57\%}}    & \multicolumn{1}{l|}{17.22\%}& \multicolumn{1}{l|}{15.30\%}               \\ \hline
\multicolumn{1}{|l|}{Swizzor.gen!I}           & \multicolumn{1}{l|}{97.57\%}        & \multicolumn{1}{l|}{98.14\%}        & \multicolumn{1}{l|}{96.80\%}        & \multicolumn{1}{l|}{\textbf{99.29\%}}    & \multicolumn{1}{l|}{33.57\%}& \multicolumn{1}{l|}{14.55\%}             \\ \hline
\multicolumn{1}{|l|}{VB.AT}                   & \multicolumn{1}{l|}{99.36\%}        & \multicolumn{1}{l|}{98.72\%}        & \multicolumn{1}{l|}{98.77\%}        & \multicolumn{1}{l|}{\textbf{99.34\%}}    & \multicolumn{1}{l|}{31.68\%}& \multicolumn{1}{l|}{13.92\%}           \\ \hline
\multicolumn{1}{|l|}{Wintrim.BX}              & \multicolumn{1}{l|}{99.78\%}        & \multicolumn{1}{l|}{97.71\%}        & \multicolumn{1}{l|}{\textbf{99.92\%}}        & \multicolumn{1}{l|}{99.88\%}    & \multicolumn{1}{l|}{31.71\%}& \multicolumn{1}{l|}{15.65\%}           \\ \hline
\multicolumn{1}{|l|}{Yuner.A}                 & \multicolumn{1}{l|}{88.26\%}        & \multicolumn{1}{l|}{84.44\%}        & \multicolumn{1}{l|}{80.64\%}        & \multicolumn{1}{l|}{\textbf{99.79\%}}    & \multicolumn{1}{l|}{16.38\%}& \multicolumn{1}{l|}{11.54\%}         \\ \hline
\end{tabular}
\end{table*}

\begin{table*} [!hbpt]
 \caption{Accuracy averages of the six tested models.}
\centering
\begin{tabular}{|c|l|l|l|l|l|l|}
\hline
\multirow{3}{*}{\textbf{Models}} & {\textbf{CNN-SVM}} & {\textbf{GRU-SVM}} & {\textbf{MLP-SVM}} & {\textbf{Inception V3}} & {\textbf{ResNet 50}} & {\textbf{VGG16}} \\ \cline{2-7} 
                        & \multicolumn{6}{c|}{\textbf{Average of prediction accuracy}}                         \\ \cline{2-7} 
                        &  \hfil 93.22\%& \hfil 94.17\%&\hfil 94.55\%&\hfil 99.25\%& \hfil 26.66\%&\hfil 14.31\%     \\ \hline
\end{tabular}
\end{table*}

\begin{table*} [!hbpt]
\centering
\setlength\tabcolsep{1pt}
\caption{Comparison of malware detection models, including models we tested.}
\begin{tabular}{|c |c |c |c |c |c |c |c |c |c |c |c |c |c |}
\hline
{\textbf{Model}}    & \rotatebox{90}{\textbf{VGG16}} & \rotatebox{90}{\textbf{ResNet50}} & \rotatebox{90}{MLP-SVM\cite{agarap2017towards}} & \rotatebox{90}{CNN-SVM\cite{agarap2017towards}} & \rotatebox{90}{GRU-SVM\cite{agarap2017towards}} & \rotatebox{90}{Random Forest\cite{garcia2016random}} & \rotatebox{90}{\textbf{MLP-SVM}} & \rotatebox{90}{\textbf{GRU-SVM}} & \rotatebox{90}{CNN\cite{kabanga2017malware}}     & \rotatebox{90}{M-CNN\cite{kalash2018malware}}    & \rotatebox{90}{\textbf{CNN-SVM}} &\rotatebox{90} {\textbf{Inception V3}} \\ \hline
{\textbf{Accuracy}} & {15.92}          & {35.10\%}           & { 80.46\%} & { 77.22\%} & { 84.92\%} & {\ 95.26\%}       & { 97.25\%}          & { 97.43\%}          & {98.00\%} & { 98.52\%} & {99.11\%}          & {\textbf{99.24\%}}      \\ \hline
\end{tabular}
\end{table*}

\section{Experimental Results}

All experiments in this study were conducted on NVIDIA GeForce GTX 1080 Ti GPU. As stated, we ran six models on the Malimg dataset:  Inception V3, VGG16-Net, ResNet50, CNN-SVM, MLP-SVM and GRU-SVM. Since the Malimg dataset is not similar to the ImageNet dataset, we could not directly use grayscale images with VGG16 and ResNet50 because the input layers require the shape of (3, 224, 224).  The 3 represents Red, Green and Blue (RGB) channels of the image, whereas the grayscale images require (1, 224, 224). VGG16 and ResNet50 showed low performance, compared to the other models, since both of these models architectures were designed to recognize colored images that requires RGB format. Therefore, both give low accuracies when tested on the grayscale images. The results for malware prediction using all these models are shown in Table 2 and Fig. 11. The Inception V3 model had a significantly higher accuracy at 99.24\%. Table 4 shows the best predicted accuracies of the six models when run 10 times.  CNN-SVM, GRU-SVM, and MLP-SVM performed well but VGG16 and ResNet50 performed poorly compared to the Inception V3 model. We provide the results of testing the dataset with several traditional models as well as other deep learning models in Table 4.

\begin{figure*} [hbt!]
\centering
	\includegraphics[width=16cm, height=10cm]{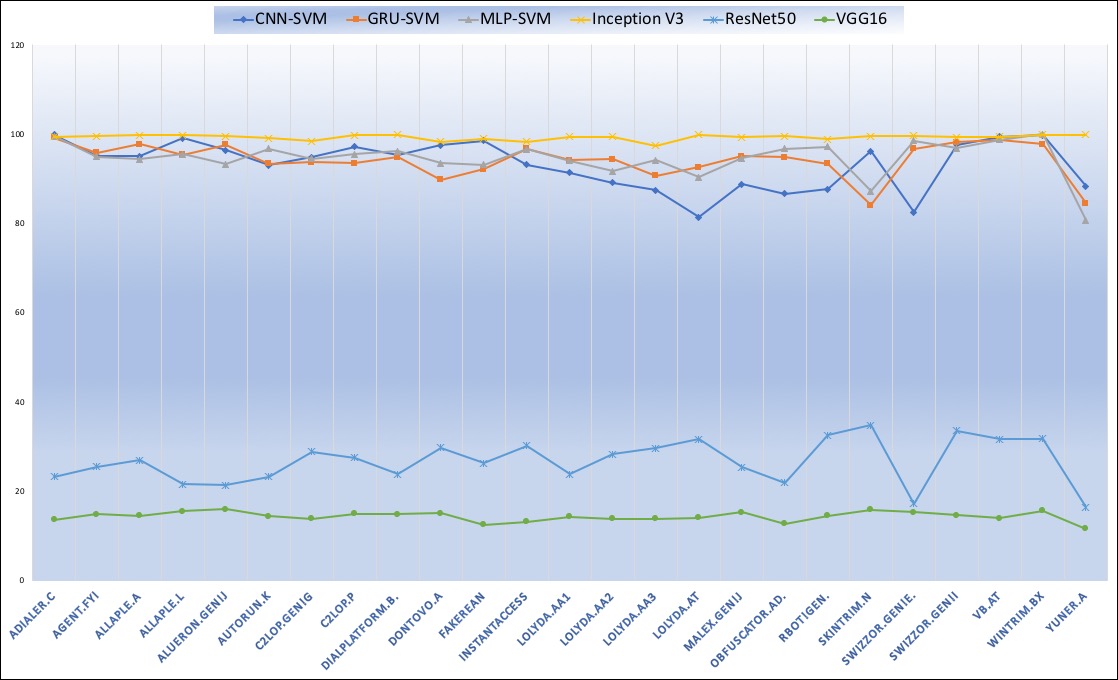}
	\caption{Prediction accuracy of six models}
\end{figure*}

\section{Conclusions and Future Work}

These days many antivirus programs rely on deep learning techniques to protect devices from malware. Deep learning architectures have achieved good performance in detecting malware when used with Windows PE binaries. We have presented the performance comparison among six classifiers on a malware image dataset created from PE files. We used the models from the ImageNet Large-Scale Visual Recognition Challenge and three other CNN models to classify grayscale malware images. We successfully trained the six models on the Malimg dataset, and the results indicate that the Inception-V3 model outperforms all compared work. To the best of our knowledge, it is the state-of-the-art of performance in classification on grayscale malware images.\\
Future work will be focused on conducting results using additional models from leaderboards of image classification competitions. We also want to convert malware images into color RGB images before classification.

\addcontentsline{toc}{chapter}{\protect\numberline{}{REFERENCES}}
\bibliography{ijns}
\bibliographystyle{unsrt}

\section*{Biography}
\noindent {\bf Ahmed Bensaoud} received B.S. degree from the Benghazi University, Libya, and M.S. from Colorado State University, Fort Collins, Colorado. Currently he is a Ph.D. student at the University of Colorado Colorado Springs. His research interests include malware detection and malware classification.\vspace*{0.2cm}\\
\noindent {\bf Nawaf Abudawaood} graduated from the University of Colorado at Colorado Springs with a Masters in Engineering in Information Assurance. He received his Bachelors degree from the Old Dominion University Norfolk, Virginia, in Information Systems and Technology. He currently works for The Exchange Hub as a Cyber Security Engineer.\vspace*{0.2cm}\\
\noindent {\bf  Jugal Kalita} received Ph.D. from the University of Pennsylvania, Philadelphia. He is a Professor of Computer Science at the University of Colorado, Colorado Springs. His research interests are in machine learning and natural language processing. He has published over 250 papers in international journals and referred conference proceedings and has written four books.

\end{document}